# Quantum teleportation of multiple degrees of freedom in a single photon

Xi-Lin Wang, Xin-Dong Cai, Zu-En Su, Ming-Cheng Chen, Dian Wu, Li Li, Nai-Le Liu, Chao-Yang Lu, and Jian-Wei Pan

Hefei National Laboratory for Physical Sciences at Microscale and Department of Modern Physics, University of Science and Technology of China,

& CAS Centre for Excellence and Synergetic Innovation Centre in Quantum Information and Quantum Physics, Hefei, Anhui 230026, China

**Quantum teleportation[1] provides a "disembodied" way to transfer quantum states from one object to another at a distant location, assisted by priorly shared entangled states and a classical communication channel. In addition to its fundamental interest, teleportation has been recognized as an important element in long-distance quantum communication[2], distributed quantum networks[3] and measurement-based quantum computation[4,5]. There have been numerous demonstrations of teleportation in different physical systems such as photons[6-8], atoms[9], ions[10,11], electrons[12], and superconducting circuits[13]. Yet, all the previous experiments were limited to teleportation of one degree of freedom (DoF) only. However, a single quantum particle can naturally possess various DoFs—internal and external—and with coherent coupling among them. A fundamental open challenge is to simultaneously teleport multiple DoFs, which is necessary to fully describe a quantum particle, thereby truly teleporting it intactly. Here, we demonstrate the first teleportation of the composite quantum states of a single photon encoded in both the spin and orbital angular momentum. We develop a method to project and discriminate hyper-entangled Bell states exploiting probabilistic quantum non-demolition measurement, which can be extended to more DoFs. We verify the teleportation for both spin-orbit product states and hybrid entangled state, and achieve a teleportation fidelity ranging from 0.57 to 0.68,**

**above the classical limit. Our work moves a step toward teleportation of more complex quantum systems, and demonstrates an enhanced capability for scalable quantum technologies.**

Quantum teleportation is a linear operation applied to the quantum states, thus teleporting multiple DoFs should be possible in theory[1]. Suppose Alice wishes to teleport to Bob the composite quantum state of a single photon (see Fig.1a), encoded in both the spin angular momentum (SAM) and the orbital angular momentum (OAM):

$$|\varphi\rangle_1 = \alpha\,|0\rangle_1^s\,|0\rangle_1^o + \beta\,|0\rangle_1^s\,|1\rangle_1^o + \gamma\,|1\rangle_1^s\,|0\rangle_1^o + \delta\,|1\rangle_1^s\,|1\rangle_1^o, \quad (1)$$

where $|0\rangle^s$ and $|1\rangle^s$ denote horizontal and vertical polarization of the SAM, and $|0\rangle^o$ and $|1\rangle^o$ refer to right-handed and left-handed OAM with $\pm\hbar$, respectively, and $\alpha$, $\beta$, $\gamma$ and $\delta$ are complex numbers satisfying $|\alpha|^2+|\beta|^2+|\gamma|^2+|\delta|^2=1$. To do so, they first need to share a hyper-entangled photon pair 2-3, which is simultaneously entangled in both the SAM and OAM:

$$|\xi\rangle_{23} = |\phi^-\rangle_{23}\,|\omega^+\rangle_{23} = \frac{1}{2}(|0\rangle_2^s\,|0\rangle_3^s - |1\rangle_2^s\,|1\rangle_3^s)(|0\rangle_2^o\,|0\rangle_3^o + |1\rangle_2^o\,|1\rangle_3^o)\,. \quad (2)$$

Here we use $|\phi^\pm\rangle = (|0\rangle^s\,|0\rangle^s \pm |1\rangle^s\,|1\rangle^s)/\sqrt{2}$ and $|\psi^\pm\rangle = (|0\rangle^s\,|1\rangle^s \pm |1\rangle^s\,|0\rangle^s)/\sqrt{2}$ to denote the four Bell states encoded with SAM, and $|\omega^\pm\rangle = (|0\rangle^o\,|0\rangle^o \pm |1\rangle^o\,|1\rangle^o)/\sqrt{2}$ and $|\chi^\pm\rangle = (|0\rangle^o\,|1\rangle^o \pm |1\rangle^o\,|0\rangle^o)/\sqrt{2}$ to denote the four Bell state in OAM. Their tensor products result in a group of 16 hyper-entangled Bell states.

A crucial step in the teleportation is to perform a two-particle joint measurement of photon 1 and 2, projecting them onto the basis of the 16 orthogonal complete hyper-entangled Bell states, and discriminating one of them, e.g.

$$|\xi\rangle_{12} = |\phi^-\rangle_{12}\,|\omega^+\rangle_{12} = \frac{1}{2}(|0\rangle_1^s\,|0\rangle_2^s - |1\rangle_1^s\,|1\rangle_2^s)(|0\rangle_1^o\,|0\rangle_2^o + |1\rangle_1^o\,|1\rangle_2^o)\,. \quad (3)$$

Such a process is referred as hyper-entangled Bell state measurement (*h*-BSM). After the *h*-BSM that projects the two photons 1 and 2 onto the state $|\xi\rangle_{12}$, quantum physics dictates that photon 3 will be projected onto the initial state of photon 1:

$$|\varphi\rangle_3 = \alpha\,|0\rangle_3^s\,|0\rangle_3^o + \beta\,|0\rangle_3^s\,|1\rangle_3^o + \gamma\,|1\rangle_3^s\,|0\rangle_3^o + \delta\,|1\rangle_3^s\,|1\rangle_3^o \;. \tag{4}$$

With an equal probability of 1/16, photons 1 and 2 can also be projected onto one of the other 15 hyper-entangled Bell states (Methods). The *h*-BSM results can be broadcast as a 4-bit classical information, which will allow Bob to apply appropriate Pauli operations to perfectly reconstruct the initial composite state of photon 1.

Experimental realization of the above teleportation protocol poses significant challenges in coherent control of multiple particles and multiple DoFs simultaneously. The most difficult task is to implement the *h*-BSM, because it would normally require coherently controlled gates between independent qubits of different DoFs. Moreover, with multiple DoFs, it is necessary to measure one DoF without disturbing another one. With linear operations only, previous theoretical work[14] has suggested that it was impossible to discriminate the hyper-entangled states unambiguously. This challenge has been overcome in our work.

Figure 1 illustrates our linear optical scheme for teleporting the spin-orbit composite state. The *h*-BSM is implemented in a step-by-step manner. First, the two photons, 1 and 2, are sent through a polarizing beam splitter (PBS), an optical device that transmits horizontal polarization ($|0\rangle^s$) and reflects vertical polarization ($|1\rangle^s$). After the PBS, we post-select the events that there is one and only one photon in each output. Such events is possible only if the two input photons have the same SAM—both are transmitted ($|0\rangle^s{}_1\,|0\rangle^s{}_2$) or reflected ($|1\rangle^s{}_1\,|1\rangle^s{}_2$)—thus projecting the SAM part of the wave function into the 2-dimensional subspace spanned by $|\phi^{\pm}\rangle = (|0\rangle^s\,|0\rangle^s \pm |1\rangle^s\,|1\rangle^s)/\sqrt{2}$. At both outputs of the PBS, we add two polarisers, projecting the two photons in the diagonal basis $(|0\rangle^s + |1\rangle^s)/\sqrt{2}$. It should be noted that the PBS is not OAM-preserving[15], as the reflection on the PBS flips the sign of the OAM qubit, that is, $|0\rangle_1^s|0\rangle_1^o \to |0\rangle_{1'}^s|0\rangle_{1'}^o$, $|0\rangle_1^s|1\rangle_1^o \to |0\rangle_{1'}^s|1\rangle_{1'}^o$, $|1\rangle_1^s|0\rangle_1^o \to i|1\rangle_{2'}^s|1\rangle_{2'}^o$, $|1\rangle_1^s|1\rangle_1^o \to i|1\rangle_{2'}^s|0\rangle_{2'}^o$. Thus, both the SAM and OAM are required to be taken into account as a molecular-like coupled state. A detailed mathematical treatment is presented in Methods, showing that the PBS and two polarizers select the 4 following states out of the total 16 hyper-entangled Bell states: $|\phi^+\rangle_{12}|\omega^-\rangle_{12}$, $|\phi^-\rangle_{12}|\omega^+\rangle_{12}$, $|\phi^-\rangle_{12}|\chi^+\rangle_{12}$, and $|\phi^-\rangle_{12}|\chi^-\rangle_{12}$.

Having measured and filtered out the SAM, secondly, we perform BSM on the remaining OAM qubit. The two single photons out of the PBS are superposed on a beam splitter (BS, see Fig.1a). It has been known that only the asymmetric Bell state will lead to a coincidence detection where there is one and only one photon in each output[16], whereas for the three other symmetric Bell state, the two input photons will coalesce to a single output mode. Note again the reflection on the BS inverts the OAM sign. Therefore, the state $|\omega^-\rangle_{12}$ can be distinguished by coincidence detection, and $|\omega^+\rangle_{12}$ can be discriminated by measuring two orthogonal OAMs in either output. In total, these two steps would allow an unambiguous discrimination of the two hyper-entangled Bell states: $|\phi^+\rangle_{12}|\omega^-\rangle_{12}$ and $|\phi^-\rangle_{12}|\omega^+\rangle_{12}$.

Unfortunately, the above two interferometers cannot be simply cascaded. The working of the interferometers rests on the assumption that two input photons are from different path. When directly connected, the 50% chance of both photons coalescing into a single output spatial mode after the first interferometer would remain undetected (no "coincidence" event is known), and would further induce erroneous coincidence event after the second interferometer, thus failing both BSMs. This is a ubiquitous and important problem in linear optical quantum information processing[5,17].

Our remedy is to exploit quantum non-demolition (QND) measurement—seeing a single photon without destroying it and keeping its quantum information intact. Interestingly, quantum teleportation itself can be used for probabilistic QND detection[7,18]. As shown in Fig.1b, another pair of photons entangled in OAM is used as ancillary. The procedure is a standard teleportation:

If there is an incoming photon, a two-photon coincidence detection behind a BS can happen with a 50% efficiency, triggering a successful BSM, which heralds the presence of the incoming photon and meanwhile faithfully teleports the quantum state of the incoming photon to a freely propagating photon. In the case of no incoming photon, two-photon coincidence behind the BS never happen, in which case we will know and ignore the out-going photon. Thus, using the QND measurement, the BSM interferometers can be concatenated. Note that using a QND in one of the arm of the BS is sufficient due to the conserved total number of eventually registered photons. Our protocol can identify two hyper-entangled Bell states with an overall efficiency of 1/32 (Methods). The QND method is also applicable to boost the efficiency of photonic quantum logic gates in the Knill-Laflamme-Milburn scheme for scalable optical quantum computing[5].

Figure 2 shows the experimental setup for the realization of quantum teleportation of the spin-orbit composite state of a single photon. Passing a femtosecond pulsed laser through three type-I β-barium borate (BBO) crystals generates three photon pairs[19-21], engineered in different forms (Methods). From the first photon pair (1-*t*) , triggered by the detection of its sister photon (*t*), a heralded single photon to be teleported is prepared in the spin-orbit composite state $|\varphi\rangle$. The second pair (2-3) is created in the hyper-entangled state $|\xi\rangle_{23}$. The third pair (4-5) is prepared as the ancillary OAM-entangled state $|\omega^{+}\rangle_{45}$ for the teleportation-based QND measurement.

To demonstrate that teleportation works universally for arbitrary quantum superposition states, we test the following five representative initial states: $|\varphi\rangle_{\mathrm{A}} = |0\rangle^{s}|0\rangle^{o}$, $|\varphi\rangle_{\mathrm{B}} = |1\rangle^{s}|1\rangle^{o}$, $|\varphi\rangle_{\mathrm{C}} = (|0\rangle^{s}+|1\rangle^{s})(|0\rangle^{o}+|1\rangle^{o})/2$, $|\varphi\rangle_{\mathrm{D}} = (|0\rangle^{s}+i|1\rangle^{s})(|0\rangle^{o}+i|1\rangle^{o})/2$, and $|\varphi\rangle_{\mathrm{E}} = (|0\rangle^{s}|0\rangle^{o}+|1\rangle^{s}|1\rangle^{o})/\sqrt{2}$. These states can be grouped into three categories. The first two, $|\varphi\rangle_{\mathrm{A}}$ and $|\varphi\rangle_{\mathrm{B}}$, are product states of the both DoFs in the computational basis. $|\varphi\rangle_{\mathrm{C}}$ and $|\varphi\rangle_{\mathrm{D}}$ are products states of both DoFs in the superposition basis. The last one, $|\varphi\rangle_{\mathrm{E}}$, is a spin-orbit hybrid entangled state. The four product states are prepared by independent single-qubit rotations, using wave plates for SAM and spiral phase plate (SPP) or binary phase plate (BPP) for OAM, respectively. The entangled state is generated by counter-propagating a SPP inside a Sagnac interferometer (Methods).

The implementation of the *h*-BSM and QND measurement requires Hong-Ou-Mandel[22] type interference of two indistinguishable single photons with good spatial and spectral overlap. To make the independent photons 1, 2, 4, and 5 indistinguishable, their arrival time is matched to be within $10\ \mathrm{fs}$, much smaller than the coherence time of the down-converted photons ($\sim 448\ \mathrm{fs}$) stretched by narrow-band spectral filtering ($\sim 3\ \mathrm{nm}$). A step-by-step verification of two-photon interference as a function of temporal delay is shown in Extended Data Fig. 1. We observe a visibility of $0.75 \pm 0.03$ for the interference of two SAM-encoded photons at the PBS, and $0.73 \pm 0.03$ ($0.69 \pm 0.03$) for the interference of two OAM-encoded photons at the $BS_1$ ($BS_2$).

The final verification of the teleportation results relies on the coincidence detection counts of the six photon encoded in both DoFs, which would suffer from a low rate. It should be noted that all the pervious experiments[15,20,21,23-26] with OAM states have never gone beyond two single photons. We overcome this challenge by preparing high-brightness hyper-entangled photons and designing dual-channel and high-efficiency OAM measurement devices (Methods).

To evaluate the performance of the teleportation, we measure the teleported state fidelity $F = Tr(\hat{\rho}\,|\varphi\rangle\langle\varphi|)$, defined as the overlap of the ideal teleported state ($|\varphi\rangle$) and the measured density matrix ($\hat{\rho}$). Conditioned on the detection of the trigger photon *t* and the four-photon coincidence after the *h*-BSM, we register the photon counts of the teleported photon 3 and analyse its composite state. The fidelity measurement for the product states are straightforward as we can measure the SAM and OAM qubits separately. We measure the final state in the $|0\rangle^s / |1\rangle^s$ and $|0\rangle^o / |1\rangle^o$ basis for the $|\varphi\rangle_{\mathrm{A}}$ and $|\varphi\rangle_{\mathrm{B}}$ teleportation, in the $(|0\rangle^s \pm |1\rangle^s)/\sqrt{2}$ and

$(|0\rangle^o \pm |1\rangle^o)/\sqrt{2}$ basis for the $|\varphi\rangle_C$ teleportation, and in the $(|0\rangle^s \pm i|1\rangle^s)/\sqrt{2}$ and $(|0\rangle^o \pm i|1\rangle^o)/\sqrt{2}$ basis for the $|\varphi\rangle_D$ teleportation, respectively. From our data shown in Fig. 3, the teleportation fidelities for $|\varphi\rangle_A$, $|\varphi\rangle_B$, $|\varphi\rangle_C$ and $|\varphi\rangle_D$ yield $0.68 \pm 0.04$, $0.66 \pm 0.04$, $0.62 \pm 0.04$ and $0.63 \pm 0.04$, respectively.

For the entangled state $|\varphi\rangle_E$ where the SAM and OAM qubits are not separable, its fidelity can be decomposed as $F_E = \frac{1}{4} Tr[\hat{\rho}(I + \hat{\sigma}_x^s \hat{\sigma}_x^o - \hat{\sigma}_y^s \hat{\sigma}_y^o + \hat{\sigma}_z^s \hat{\sigma}_z^o)]$, where $\sigma_x$, $\sigma_y$ and $\sigma_z$ are the Pauli operators. The expectation values of the joint observables, $\hat{\sigma}_x^s \hat{\sigma}_x^o$, $\hat{\sigma}_y^s \hat{\sigma}_y^o$ and $\hat{\sigma}_z^s \hat{\sigma}_z^o$, can be obtained by local measurements in the corresponding basis for the two DoFs, $(|0\rangle \pm |1\rangle)/\sqrt{2}$, $(|0\rangle \pm i|1\rangle)/\sqrt{2}$, and $|0\rangle/|1\rangle$, respectively. Our experimental results on the three different basis are presented in Fig. 4a-c, from which we determine a teleportation fidelity of $0.57 \pm 0.02$ for the entangled state.

We note that all reported data are without background subtraction. The main sources of error include double pair emission, imperfection in the initial states, entangled photons 2-3 and 4-5, two-photon interferences, and OAM measurement. We note that the teleportation fidelity of the states in the three categories are affected differently by the error sources (Methods). The measured fidelities of the five typical teleported states (summarised in Fig. 4d) are all well beyond 0.40—the classical limit, defined as the optimal state estimation fidelity on a single copy of a two-qubit system[27]. These results prove the successful realization of quantum teleportation of the spin-orbit composite state of a single photon. Furthermore, for the entangled state $|\varphi\rangle_E$, we emphasize that the teleportation fidelity exceeds the threshold of 0.5 for proving the presence of entanglement[28], which demonstrates that the hybrid entanglement of different DoFs inside a quantum particle can preserve after the teleportation.

In summary, we present the first quantum teleportation of multiple properties of a single quantum particle, demonstrating the ability to coherently control and simultaneously teleport a single object with multiple DoFs that can form a hybrid entangled state. It is interesting to note that these DoFs can also be in a fully undefined state, such as being part of a hyper-entangled pair, which would lead to the protocol of hyper-entanglement swapping[29]. Our methods can be generalized to more DoFs (see Methods for a universal scheme). The efficiency of teleportation, which is limited mainly by the efficiency of $h$-BSM (1/32 in the present experiment), can be enhanced by using more ancillary photons, quantum encoding, embedded teleportation tricks, high-efficiency single-photon detectors, and active feed-forward, in a spirit similar to the Knill-Laflamme-Milburn scheme[5]. We didn't implement the feed-forward in the current experiment, which could be done using electric-optical modulators for both the SAM and OAM qubits (see Methods for a detailed protocol). Although our current work is based on linear optics and single photons, the multi-DoF teleportation protocol is by no means limited to this system, but can also be applied to other quantum systems such as trapped electrons[12], atoms[9], and ions[10,11].

Besides the fundamental interest, the developed methods in this work on the manipulation of quantum states of multiple DoFs will open up new possibilities in quantum technologies. Controlling multiple DoFs can bring into new advantages, such as complete SAM Bell-state analysis[23] and alignment-free quantum communication[30]. Carrying multiple DoFs on photons can enhance the information capacity in quantum communication protocols such as quantum super-dense coding[23]. Moreover, combining the entanglement of multiple ($M$-) photons and high ($N$-) dimensions from the multiple DoFs would allow the generation of hyper-entanglement with an expanded Hilbert space size that grows as $N^M$, which would provide a versatile platform in the near future for demonstrations of complex quantum communication and quantum computing protocols, as well as extreme violations of Bell inequalities.

## References

1. Bennett, C. H. *et al.* Teleporting an unknown quantum state via dual classic and Einstein-Podolsky-Rosen channels. *Phys. Rev. Lett.* **70**, 1895–1899 (1993).

2. Briegel, H. J., Dur, W., Cirac, J. I. & Zoller, P. Quantum repeaters: the role of imperfect local operations in quantum communication. *Phys. Rev. Lett.* **81**, 5932–5935 (1998).
3. Kimble, H. J. The quantum internet. *Nature* **453**, 1023–1030 (2008).
4. Gottesman, D. & Chuang, I. Demonstrating the viability of universal quantum computation using teleportation and single-qubit operations. *Nature* **402**, 390–393 (1999).
5. Knill, E., Laflamme, R. & Milburn, G. J. A scheme for efficient quantum computation with linear optics. *Nature* **409**, 46–52 (2001).
6. Bouwmeester, D. *et al.* Experimental quantum teleportation. *Nature* **390**, 575–579 (1997).
7. Marcikic, I., de Riedmatten, H., Tittel, W., Zbinden, H. & Gisin, N. Long-distance teleportation of qubits at telecommunication wavelengths. *Nature* **421**, 509–513 (2003).
8. Takeda, S., Mizuta, T., Fuwa, M., van Loock, P. & Furusawa, A. Deterministic quantum teleportation of photonic quantum bits by a hybrid technique. *Nature* **500**, 315–318 (2013).
9. Krauter, H. *et al.* Deterministic quantum teleportation between distant atomic objects. *Nat. Phys.* **9**, 400–404 (2013).
10. Riebe, M. *et al.* Deterministic quantum teleportation with atoms. *Nature* **429**, 734–737 (2004).
11. Barrett, M. D. *et al.* Deterministic quantum teleportation of atomic qubits. *Nature* **429**, 737–739 (2004).
12. Pfaff, W. *et al.* Unconditional quantum teleportation between distant solid-state quantum bits. *Science* **345**, 532–535 (2014).
13. Steffen, L. *et al.* Deterministic quantum teleportation with feed-forward in a solid state system. *Nature* **500**, 319–322 (2013).
14. Wei, T. C., Barreiro, J. T. & Kwiat, P. G. Hyperentangled Bell-state analysis. *Phys. Rev. A* **75**, 060305(R) (2007).
15. Nagali, E. *et al.* Optimal quantum cloning of orbital angular momentum photon qubits through Hong–Ou–Mandel coalescence. *Nature Photon.* **3**, 720–723 (2009).

16. Weinfurter, H. Experimental Bell-state analysis. *Europhys. Lett.* **25**, 559–564 (1994).
17. Pan, J.-W. *et al*. Multiphoton entanglement and interferometry. *Rev. Mod. Phys.* **84**, 777–838 (2012).
18. B. C. Jacobs, T. B. Pittman & J. D. Franson, Quantum relays and noise suppression using linear optics. Phys. Rev. A 66, 052307 (2002).
19. Kwiat, P. G. *et al*. Ultrabright source of polarization-entangled photons. *Phys. Rev. A* **60**, 773-776 (R) (1999).
20. Barreiro, J. T., Langford, N. K., Peters, N. A. & Kwiat, P. G. Generation of hyper-entangled photon pairs. *Phys. Rev. Lett*. **95**, 260501 (2005).
21. Mair, A., Vaziri, A., Weihs, G. & Zeilinger, A. Entanglement of the orbital angular momentum states of photons. *Nature* **412**, 313–316 (2001).
22. Hong, C. K.; Ou, Z. Y. & Mandel, L. Measurement of subpicosecond time intervals between two photons by interference. *Phys. Rev. Lett.* **59**, 2044–2046 (1987).
23. Barreiro, J. T., Wei, T.-C. & Kwiat, P. G. Beating the channel capacity limit for linear photonic superdense coding. *Nature Phys.* **4**, 282–286 (2008).
24. Nagali, E. *et al*. Quantum information transfer from spin to orbital angular momentum of photons. *Phys. Rev. Lett*. **103**, 013601 (2009).
25. Leach, J. *et al.* Quantum correlations in optical angle-orbital angular momentum variables. *Science* **329**, 662–665 (2010).
26. Fickler, R. *et al.* Quantum entanglement of high angular momenta. *Science* **338**, 640–643 (2012).
27. Hayashi, A., Hashimoto, T. & Horibe, M. Reexamination of optimal quantum state estimation of pure states. *Phys. Rev. A* 72, 032325 (2005).
28. Gühne, O. & Toth, G. Entanglement detection. *Phys. Rep.* **474**, 1–75 (2009).
29. Zukowski, M., Zeilinger, A., Horne, M. A. & Ekert, A. "Event-ready-detectors" Bell experiment via entanglement swapping. *Phys. Rev. Lett.* **71**, 4287–4290 (1993).

30. D'Ambrosio, V. *et al*. Complete experimental toolbox for alignment-free quantum communication. *Nature Commun*. **3**, 961 (2012).

**Acknowledgements:** This work was supported by the National Natural Science Foundation of China, the Chinese Academy of Sciences, and the National Fundamental Research Program (Grant No. 2011CB921300).

**Author contributions:** C.-Y.L. and J.-W.P. conceived and designed the research, X.-L.W., X.-D.C., Z.-E.S., M.-C.C., D.W., L.L., N.-L.L., and C.-Y.L. performed the experiment, X.-L.W., M.-C.C., C.-Y.L. and J.-W.P. analyzed the data, C.-Y.L. and J.-W.P. wrote the paper with input from all authors, N.-L.L., C.-Y.L. and J.-W.P. supervised the whole project.

The authors declare no competing financial interests.

Correspondence and requests for materials should be addressed to C.-Y.L. (cylu@ustc.edu.cn) or J.-W.P. (pan@ustc.edu.cn).

## Figure captions

**Figure 1 | Scheme for quantum teleportation of the spin-orbit composite states of a single photon**. **a,** Alice wishes to teleport to Bob the quantum state of a single photon 1 which possess multiple DoFs. The quantum state of the photon 1 is encoded in both its SAM and OAM. To do so, Alice and Bob need to share a hyper-entangled photon pair 2-3. Alice then carries out an *h*-BSM assisted by a QND measurement (see main text for details) and tell the results as a 4-bit classical information to Bob. On receiving Alice's *h*-BSM result, Bob can apply appropriate Pauli operations on photon 3 to convert it into the original state of photon 1. The active feed-forward is essential for a full, deterministic teleportation. In our present proof-of-principle experiment, we didn't apply feed-forward but used post-selection to verify the success of teleportaiton. **b,** Teleportation-based probabilistic QND measurement with an ancillary entangled photon pair. An incoming photon can cause a coincidence detection after the BS, which heralds its presence and meanwhile intactly teleports its arbitrary unknown quantum state to a free-flying photon. In the case of no incoming photon, the coincidence after the BS cannot happen, thus knowing its absence.

**Figure 2 | Experimental setup for teleporting multiple properties of a single photon**. A pulsed ultraviolet (UV) laser is focused on three β-barium borate (BBO) crystals and produces three photon pairs in spatial modes 1–*t*, 2–3 and 4–5. Triggered by its sister photon *t*, photon 1 is initialised in various spin-orbit composite states ($|\varphi\rangle_{\mathrm{A},\cdots\mathrm{E}}$) to be teleported. The second pair 2-3 is hyper-entangled in both SAM and OAM. The third pair 4-5 is OAM-entangled. The *h*-BSM for the photons 1 and 2 are performed in three steps: (1) SAM BSM; (2) QND measurement; (3) OAM BSM. The teleported state is measured separately in SAM and OAM: a PBS, a HWP and a QWP are combined for SAM qubit analysis, and a SPP or a BPP together with a single-mode fibre are used for OAM qubit analysis.

**Figure 3 | Experimental results for quantum teleportation of spin-orbit product state $|\varphi\rangle_{\mathrm{A}}$, $|\varphi\rangle_{\mathrm{B}}$, $|\varphi\rangle_{\mathrm{C}}$ and $|\varphi\rangle_{\mathrm{D}}$ of a single photon**. **a-b**, Measurement result of the final state of the teleported photon 3 in the $|0\rangle^{s}/|1\rangle^{s}$ and $|0\rangle^{o}/|1\rangle^{o}$ basis for the $|\varphi\rangle_{\mathrm{A}}$ and $|\varphi\rangle_{\mathrm{B}}$ teleportation experiment. **c**, The results for the $|\varphi\rangle_{\mathrm{C}}$ teleportation, measured in the $(|0\rangle^{s}\pm|1\rangle^{s})/\sqrt{2}$ and $(|0\rangle^{o}\pm|1\rangle^{o})/\sqrt{2}$ basis. **d**, The results for the $|\varphi\rangle_{\mathrm{D}}$ teleportation, measured in the $(|0\rangle^{s}\pm i|1\rangle^{s})/\sqrt{2}$ and $(|0\rangle^{o}\pm i|1\rangle^{o})/\sqrt{2}$ basis. The X axis uses Pauli denotation for both DoFs: $|0\rangle_{z}/|1\rangle_{z}=|0\rangle/|1\rangle$, $|0\rangle_{x}/|1\rangle_{x}=(|0\rangle\pm|1\rangle)/\sqrt{2}$, and $|0\rangle_{y}/|1\rangle_{y}=(|0\rangle\pm i|1\rangle)/\sqrt{2}$. The Y axis is six-photon coincidence counts. The error bar is one standard deviation, calculated from Poissonian counting statistics of the raw detection events.

**Figure 4 | Experimental results for quantum teleportation of spin-orbit entanglement of a single photon**. To determine the state fidelity of the teleported entangled state $|\varphi\rangle_{\mathrm{E}}=(|0\rangle^{s}|0\rangle^{o}+|1\rangle^{s}|1\rangle^{o})/\sqrt{2}$, three measurement basis are required: **a**, $|0\rangle^{s}/|1\rangle^{s}$ and $|0\rangle^{o}/|1\rangle^{o}$, **b**, $(|0\rangle^{s}\pm|1\rangle^{s})/\sqrt{2}$ and $(|0\rangle^{o}\pm|1\rangle^{o})/\sqrt{2}$, and **c** $(|0\rangle^{s}\pm i|1\rangle^{s})/\sqrt{2}$ and $(|0\rangle^{o}\pm i|1\rangle^{o})/\sqrt{2}$, which are used to extract the expectation values of the joint Pauli observables $\hat{\sigma}_{z}^{s}\hat{\sigma}_{z}^{o}$, $\hat{\sigma}_{x}^{s}\hat{\sigma}_{x}^{o}$, and $\hat{\sigma}_{y}^{s}\hat{\sigma}_{y}^{o}$, respectively. Each measurement takes 12 h. The data set in **a** determines the population of the two desired terms: $|0\rangle^{s}|0\rangle^{o}$ and $|1\rangle^{s}|1\rangle^{o}$ in the entangled state $|\varphi\rangle_{\mathrm{E}}$. The data set in **b** and **c** measured in the superposition basis determines the coherence of the entangled state. We use the same Pauli denotation as in Fig. 3. **d**, A summary of the teleportation fidelities for the states $|\varphi\rangle_{\mathrm{A},\cdots\mathrm{E}}$. The error bars represent one standard deviation, deduced from propagated Poissonian counting statistics of the raw detection events.

**Figures:**

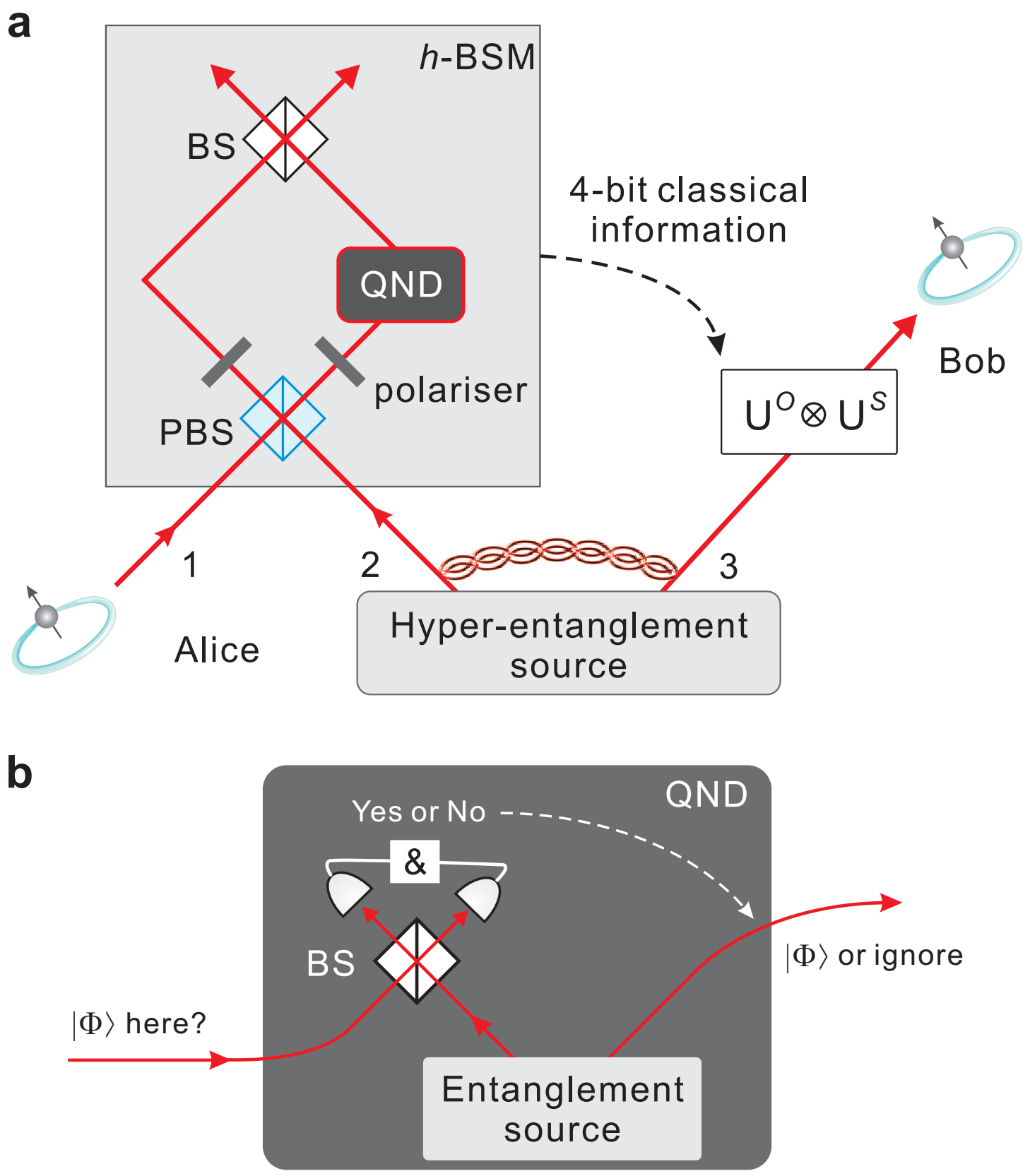

Figure 1

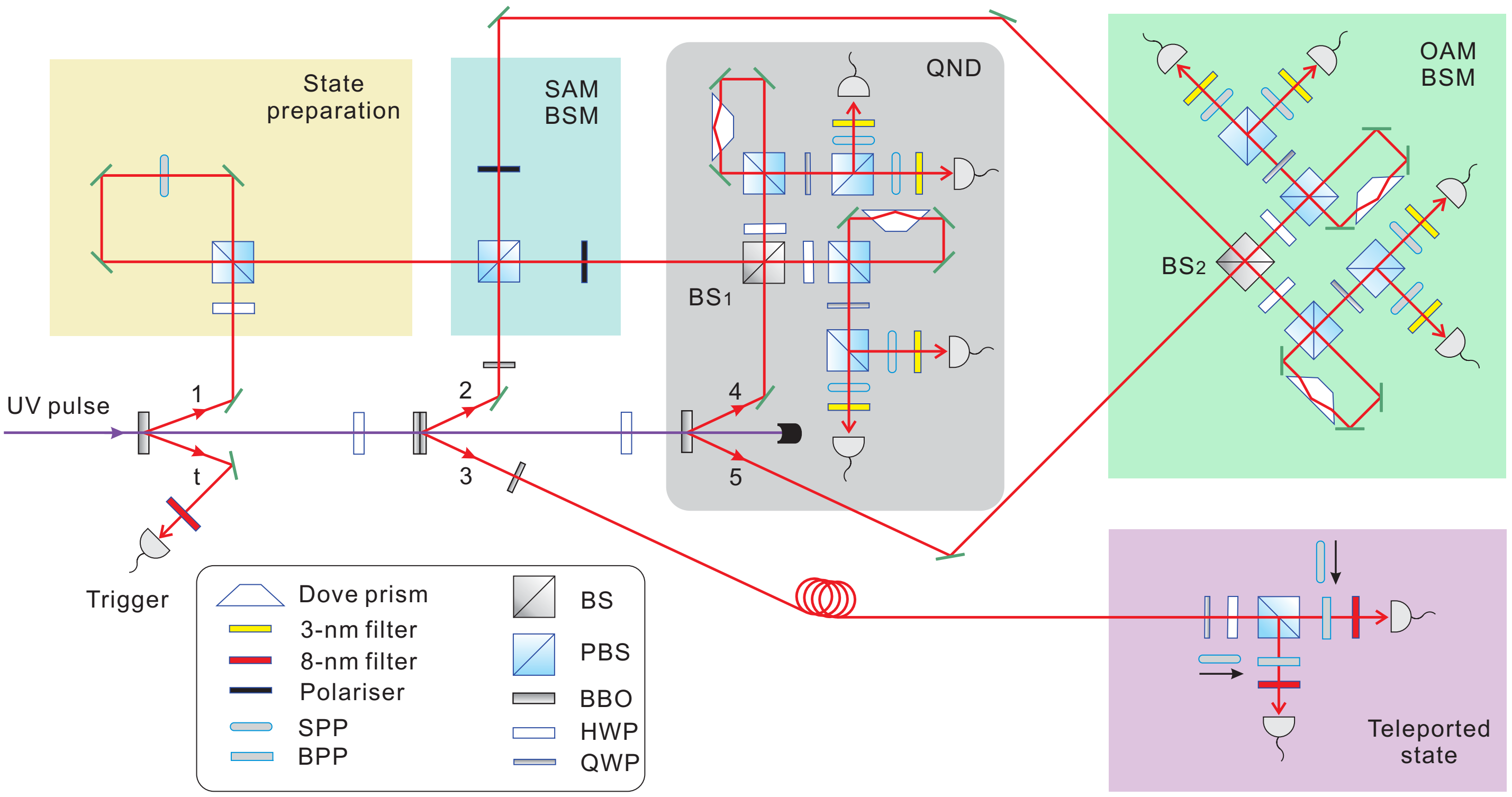

Figure 2

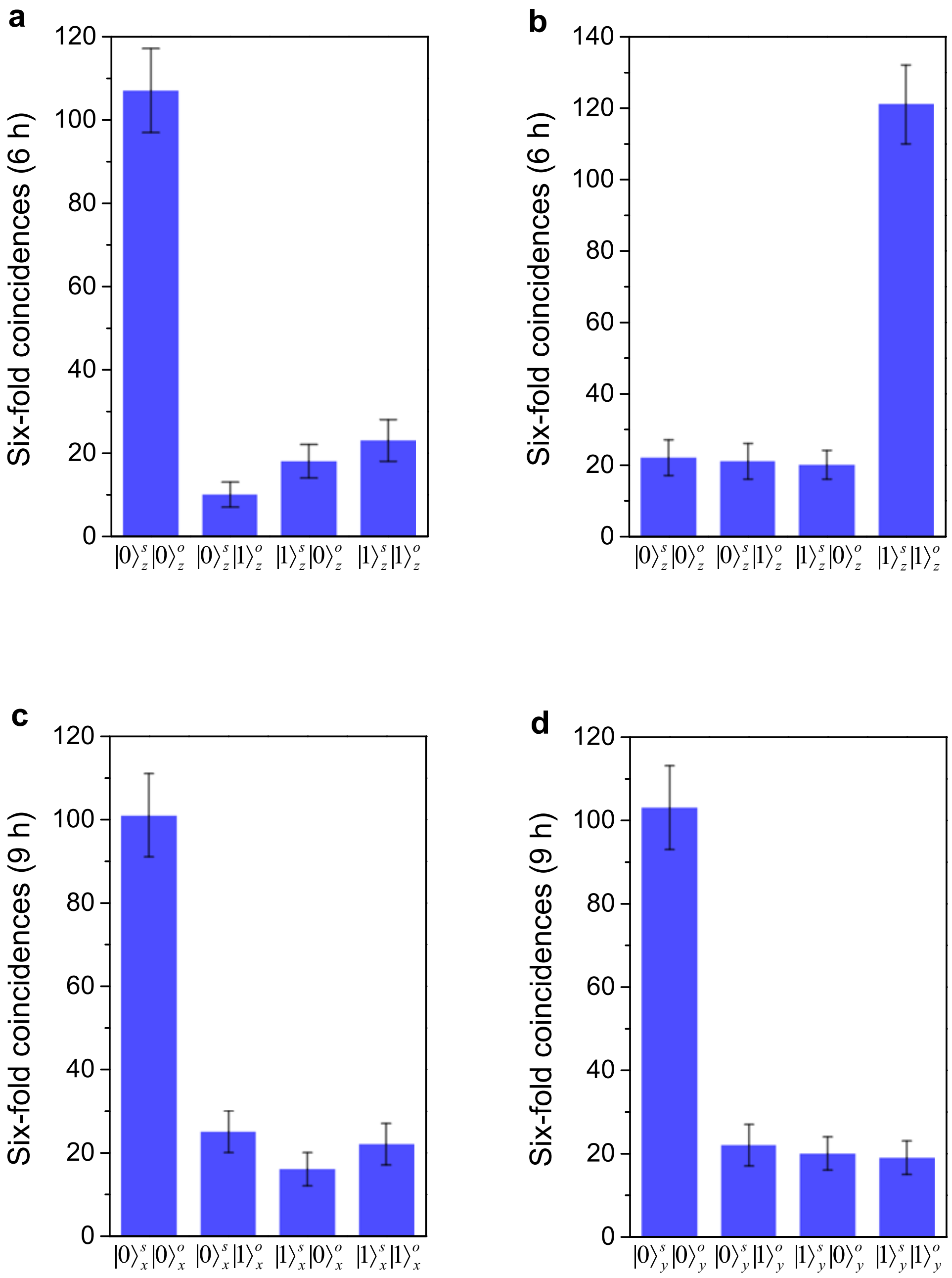

Figure 3

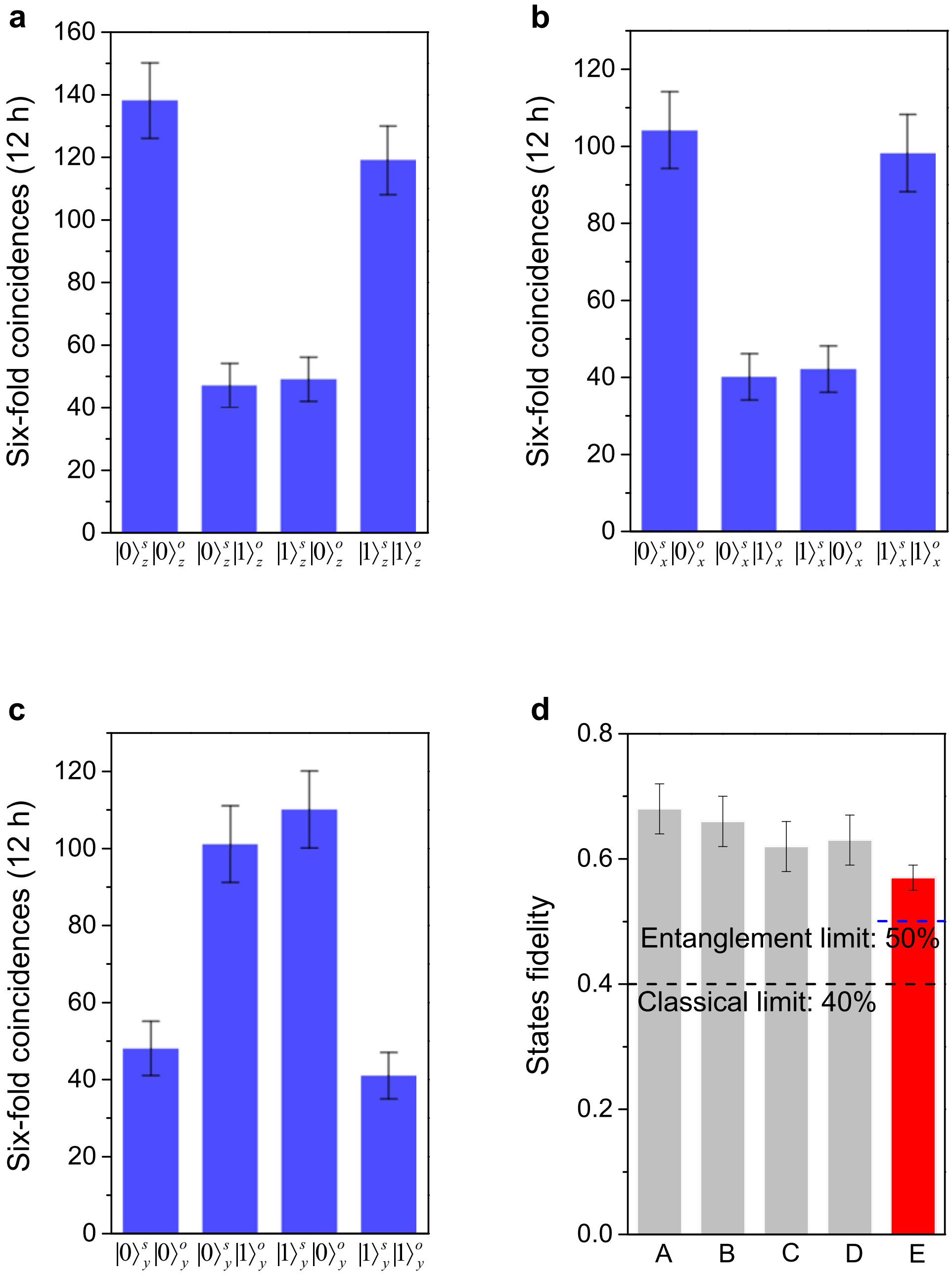
a
Six-fold coincidences (12 h)
$|0\rangle_z^s|0\rangle_z^o$ $|0\rangle_z^s|1\rangle_z^o$ $|1\rangle_z^s|0\rangle_z^o$ $|1\rangle_z^s|1\rangle_z^o$
b
Six-fold coincidences (12 h)
$|0\rangle_x^s|0\rangle_x^o$ $|0\rangle_x^s|1\rangle_x^o$ $|1\rangle_x^s|0\rangle_x^o$ $|1\rangle_x^s|1\rangle_x^o$
c
Six-fold coincidences (12 h)
$|0\rangle_y^s|0\rangle_y^o$ $|0\rangle_y^s|1\rangle_y^o$ $|1\rangle_y^s|0\rangle_y^o$ $|1\rangle_y^s|1\rangle_y^o$
d
States fidelity
Entanglement limit: 50%
Classical limit: 40%
A B C D E

Figure 4

**Methods:**

### The protocol for teleporting a spin-orbit composite quantum state

The combined state of photon 1, 2, and 3 can be rewritten in the basis of the 16 orthogonal complete hyper-entangled Bell states as following:

$$
\begin{aligned}
|\varphi\rangle_1|\xi\rangle_{23} = \frac{1}{4}[&(|\phi^+\rangle_{12}|\omega^+\rangle_{12})\hat{\sigma}^s_{3z}|\varphi\rangle_3 + (|\phi^+\rangle_{12}|\omega^-\rangle_{12})\hat{\sigma}^s_{3z}\hat{\sigma}^o_{3z}|\varphi\rangle_3 \\
&+(|\phi^+\rangle_{12}|\chi^+\rangle_{12})\hat{\sigma}^s_{3z}\hat{\sigma}^o_{3x}|\varphi\rangle_3 - (|\phi^+\rangle_{12}|\chi^-\rangle_{12})\mathrm{i}\,\hat{\sigma}^s_{3z}\hat{\sigma}^o_{3y}|\varphi\rangle_3 + (|\phi^-\rangle_{12}|\omega^+\rangle_{12})|\varphi\rangle_3 \\
&+(|\phi^-\rangle_{12}|\omega^-\rangle_{12})\hat{\sigma}^o_{3z}|\varphi\rangle_3 + (|\phi^-\rangle_{12}|\chi^+\rangle_{12})\hat{\sigma}^o_{3x}|\varphi\rangle_3 - (|\phi^-\rangle_{12}|\chi^-\rangle_{12})\mathrm{i}\,\hat{\sigma}^o_{3y}|\varphi\rangle_3 \\
&+(|\psi^+\rangle_{12}|\omega^+\rangle_{12})\mathrm{i}\,\hat{\sigma}^s_{3y}|\varphi\rangle_3 + (|\psi^+\rangle_{12}|\omega^-\rangle_{12})\mathrm{i}\,\hat{\sigma}^s_{3y}\hat{\sigma}^o_{3z}|\varphi\rangle_3 + (|\psi^+\rangle_{12}|\chi^+\rangle_{12})\mathrm{i}\,\hat{\sigma}^s_{3y}\hat{\sigma}^o_{3x}|\varphi\rangle_3 \\
&+(|\psi^+\rangle_{12}|\chi^-\rangle_{12})\hat{\sigma}^s_{3y}\hat{\sigma}^o_{3y}|\varphi\rangle_3 - (|\psi^-\rangle_{12}|\omega^+\rangle_{12})\hat{\sigma}^s_{3x}|\varphi\rangle_3 - (|\psi^-\rangle_{12}|\omega^-\rangle_{12})\hat{\sigma}^s_{3x}\hat{\sigma}^o_{3z}|\varphi\rangle_3 \\
&-(|\psi^-\rangle_{12}|\chi^+\rangle_{12})\hat{\sigma}^s_{3x}\hat{\sigma}^o_{3x}|\varphi\rangle_3 + (|\psi^-\rangle_{12}|\chi^-\rangle_{12})\mathrm{i}\,\hat{\sigma}^s_{3x}\hat{\sigma}^o_{3y}|\varphi\rangle_3]
\end{aligned}
$$

where $\hat{\sigma}^s_x$, $\hat{\sigma}^s_y$ and $\hat{\sigma}^s_z$ ($\hat{\sigma}^o_x$, $\hat{\sigma}^o_y$ and $\hat{\sigma}^o_z$) are the Pauli operators for SAM (OAM) qubits. It indicates that, regardless of the unknown sate $|\varphi\rangle_1$, the 16 measurement outcomes are equally likely, each with a probability of 1/16. By carrying out the hyper-entangled Bell state measurement (*h*-BSM) on photons 1 and 2 to unambitiously distinguish one from the group of 16 hyper-entangled Bell states, Alice can project photon 3 onto one of the 16 corresponding states. After Alice tells Bob her *h*-BSM result as a 4-bit classical information via a classical communication channel, Bob can convert the state of his photon 3 into the original state by applying appropriate two-qubit local unitary transforms $[(\hat{\mathrm{I}},\ \hat{\sigma}^s_x,\ \hat{\sigma}^s_y,\ \hat{\sigma}^s_z)\otimes(\hat{\mathrm{I}},\ \hat{\sigma}^o_x,\ \hat{\sigma}^o_y,\ \hat{\sigma}^o_z)]$.

### Two-photon interference of spin-orbit composite state on a PBS

The input and output of two photons 1 and 2 encoded in SAM and OAM on a polarizing beam splitter (PBS) as shown in Fig.1 can be summarised as follows: $|0\rangle^s_1|0\rangle^o_1 \to |0\rangle^s_{1'}|0\rangle^o_{1'}$, $|0\rangle^s_1|1\rangle^o_1 \to |0\rangle^s_{1'}|1\rangle^o_{1'}$, $|1\rangle^s_1|0\rangle^o_1 \to i|1\rangle^s_{2'}|1\rangle^o_{2'}$, $|1\rangle^s_1|1\rangle^o_1 \to i|1\rangle^s_{2'}|0\rangle^o_{2'}$. Note that PBS is not OAM-preserving, as the reflection flips the sign of OAM. Therefore, the output state for each of the input 16 hyper-entangled Bell states can be listed below:

$$|\phi^-\rangle_{12}|\omega^+\rangle_{12} \rightarrow \frac{1}{\sqrt{2}}(D_{1'}D_{2'}+A_{1'}A_{2'})|\omega^+\rangle_{1'2'}\,; \tag{S1.1}$$

$$|\phi^+\rangle_{12}|\omega^-\rangle_{12} \rightarrow \frac{1}{\sqrt{2}}(D_{1'}D_{2'}+A_{1'}A_{2'})|\omega^-\rangle_{1'2'}\,; \tag{S1.2}$$

$$|\phi^-\rangle_{12}|\chi^+\rangle_{12} \rightarrow \frac{1}{\sqrt{2}}(D_{1'}D_{2'}+A_{1'}A_{2'})|\chi^+\rangle_{1'2'}\,; \tag{S1.3}$$

$$|\phi^-\rangle_{12}|\chi^-\rangle_{12} \rightarrow \frac{1}{\sqrt{2}}(D_{1'}D_{2'}+A_{1'}A_{2'})|\chi^-\rangle_{1'2'}\,; \tag{S1.4}$$

$$|\phi^+\rangle_{12}|\omega^+\rangle_{12} \rightarrow \frac{1}{\sqrt{2}}(D_{1'}A_{2'}+A_{1'}D_{2'})|\omega^+\rangle_{1'2'}\,; \tag{S1.5}$$

$$|\phi^-\rangle_{12}|\omega^-\rangle_{12} \rightarrow \frac{1}{\sqrt{2}}(D_{1'}A_{2'}+A_{1'}D_{2'})|\omega^-\rangle_{1'2'}\,; \tag{S1.6}$$

$$|\phi^+\rangle_{12}|\chi^+\rangle_{12} \rightarrow \frac{1}{\sqrt{2}}(D_{1'}A_{2'}+A_{1'}D_{2'})|\chi^+\rangle_{1'2'}\,; \tag{S1.7}$$

$$|\phi^+\rangle_{12}|\chi^-\rangle_{12} \rightarrow \frac{1}{\sqrt{2}}(D_{1'}A_{2'}+A_{1'}D_{2'})|\chi^-\rangle_{1'2'}\,; \tag{S1.8}$$

$$|\psi^+\rangle_{12}|\omega^+\rangle_{12} \rightarrow \frac{i}{2}(D_{1'}r_{1'}D_{1'}l_{1'}-A_{1'}r_{1'}A_{1'}l_{1'}+D_{2'}r_{2'}D_{2'}l_{2'}-A_{2'}r_{2'}A_{2'}l_{2'})\,; \tag{S1.9}$$

$$|\psi^-\rangle_{12}|\omega^+\rangle_{12} \rightarrow \frac{i}{2}(D_{1'}r_{1'}D_{1'}l_{1'}-A_{1'}r_{1'}A_{1'}l_{1'}-D_{2'}r_{2'}D_{2'}l_{2'}+A_{2'}r_{2'}A_{2'}l_{2'})\,; \tag{S1.10}$$

$$|\psi^+\rangle_{12}|\omega^-\rangle_{12} \rightarrow \frac{i}{2}(D_{1'}r_{1'}A_{1'}l_{1'}-A_{1'}r_{1'}D_{1'}l_{1'}+D_{2'}r_{2'}A_{2'}l_{2'}-A_{2'}r_{2'}D_{2'}l_{2'})\,; \tag{S1.11}$$

$$|\psi^-\rangle_{12}|\omega^-\rangle_{12} \rightarrow \frac{i}{2}(D_{1'}r_{1'}A_{1'}l_{1'}-A_{1'}r_{1'}D_{1'}l_{1'}-D_{2'}r_{2'}A_{2'}l_{2'}+A_{2'}r_{2'}D_{2'}l_{2'})\,; \tag{S1.12}$$

$$|\psi^+\rangle_{12}|\chi^+\rangle_{12} \rightarrow \frac{i}{4}[(D_{1'}D_{1'}-A_{1'}A_{1'})(r_{1'}r_{1'}+l_{1'}l_{1'})+(D_{2'}D_{2'}-A_{2'}A_{2'})(r_{2'}r_{2'}+l_{2'}l_{2'})]\,; \tag{S1.13}$$

$$|\psi^-\rangle_{12}|\chi^+\rangle_{12} \rightarrow \frac{i}{4}[(D_{1'}D_{1'}-A_{1'}A_{1'})(r_{1'}r_{1'}+l_{1'}l_{1'})-(D_{2'}D_{2'}-A_{2'}A_{2'})(r_{2'}r_{2'}+l_{2'}l_{2'})]\,; \tag{S1.14}$$

$$|\psi^+\rangle_{12}|\chi^-\rangle_{12} \rightarrow \frac{i}{4}[(D_{1'}D_{1'}-A_{1'}A_{1'})(r_{1'}r_{1'}-l_{1'}l_{1'})-(D_{2'}D_{2'}-A_{2'}A_{2'})(r_{2'}r_{2'}-l_{2'}l_{2'})]\,; \tag{S1.15}$$

$$|\psi^-\rangle_{12}|\chi^-\rangle_{12} \rightarrow \frac{i}{4}[(D_{1'}D_{1'}-A_{1'}A_{1'})(r_{1'}r_{1'}-l_{1'}l_{1'})+(D_{2'}D_{2'}-A_{2'}A_{2'})(r_{2'}r_{2'}-l_{2'}l_{2'})]\,, \tag{S1.16}$$

where $D = (|0\rangle^s + |1\rangle^s)/\sqrt{2}$ , $A = (|0\rangle^s - |1\rangle^s)/\sqrt{2}$ , $r = |0\rangle^o$ , $l = |1\rangle^o$ . Then both the output modes are passing through two polarisers at $45^\circ$. Only the output with the term $D_{1'}D_{2'}$ will result in one photon in each output mode with an efficiency of 1/8. The other terms will be rejected by the conditional detection. According to (S.1.1)-(S.1.16), 4 out of the 16 output modes have the term $D_{1'}D_{2'}$ and the corresponding 4 input hyper-entangled Bell states are $|\phi^+\rangle_{12}|\omega^-\rangle_{12}, |\phi^-\rangle_{12}|\omega^+\rangle_{12}, |\phi^-\rangle_{12}|\chi^+\rangle_{12}, |\phi^-\rangle_{12}|\chi^-\rangle_{12}$, which will be sent to the next stage of BSM on the OAM qubits.

**BSM and teleportation of OAM qubits**

Having measured and filtered out the SAM qubit, next we perform BSM on the OAM qubit. Dealing with a single degree of freedom is more straightforward, which can be implemented by a beam splitter (BS). Similar to PBS, BS is also not OAM-preserving, as the reflection on the BS will also flip the sign of OAM. Therefore, the transform rules on a BS for the four OAM Bell states are:

$$|\omega^+\rangle_{12} \to i(|0\rangle_{1'}^o|1\rangle_{1'}^o + |0\rangle_{2'}^o|1\rangle_{2'}^o)/\sqrt{2}\,; \qquad \text{(S2.1)}$$

$$|\omega^-\rangle_{12} \to (|0\rangle_{1'}^o|0\rangle_{2'}^o - |1\rangle_{1'}^o|1\rangle_{2'}^o)/\sqrt{2}\,; \qquad \text{(S2.2)}$$

$$|\chi^+\rangle_{12} \to i(|0\rangle_{1'}^o|0\rangle_{1'}^o + |1\rangle_{1'}^o|1\rangle_{1'}^o + |0\rangle_{2'}^o|0\rangle_{2'}^o + |1\rangle_{2'}^o|1\rangle_{2'}^o)/2\sqrt{2}\,; \qquad \text{(S2.3)}$$

$$|\chi^-\rangle_{12} \to i(|0\rangle_{1'}^o|0\rangle_{1'}^o - |1\rangle_{1'}^o|1\rangle_{1'}^o - |0\rangle_{2'}^o|0\rangle_{2'}^o + |1\rangle_{2'}^o|1\rangle_{2'}^o)/2\sqrt{2}\,. \qquad \text{(S2.4)}$$

We can see that only the Bell state $|\omega^-\rangle_{12}$ will result in one and only one photon in each output, whereas for the three other Bell states, the two input photons will coalesce into a single output mode. Among these three Bell states, the state $|\omega^+\rangle_{12}$ can be further distinguished by measuring two single photons in either output with OAM orthogonal basis $|0\rangle^o/|1\rangle^o$. Thus, this would allow us to unambiguously discriminate 2 from the 4 Bell states. Experimentally, we design dual-channel OAM readout devices to measure the $|0\rangle^o$ and $|1\rangle^o$ simultaneously. Therefore, the efficiency for both QND and BSM on OAM is 1/2. The overall efficiency of *h*-BSM combing all the three steps is $1/8\times1/2\times1/2=1/32$.

**Generating three photon pairs**:

Ultrafast laser pulses with an average power of 800 mW, central wavelength of 394 nm, pulse duration of 120 fs, and repetition rate of 76 MHz, successively pass through three BBO crystals (Fig. 2) to generate three photon pairs through type-I spontaneous parametric down-conversion (SPDC). The first and the third crystal are 2-mm thick BBOs, whereas the second one consists of two 0.6-mm thick contiguous type-I BBO crystals with optic axes aligned in perpendicular planes. All the down converted photons have central wavelength at 788 nm. The first photon pair 1-*t* is initially prepared in the zero-order OAM mode with a coincidence count rate of $3.6\times10^5$ Hz. In SPDC, the zero-order OAM mode occupies the highest weight among all OAM modes and thus have highest brightness. The second photon pair 2-3 is simultaneously entangled in both the SAM and OAM (in the first-order mode), due to OAM conservation in the type-I SPDC. It has a count rate of $3.4\times10^4$ Hz and a state fidelity of 0.95. The third photon pair 4-5 is created in the first-order OAM entangled state with a two-photon count rate of $1.2\times10^5$ Hz, and a fidelity of 0.91. We estimate the mean number of photon pairs generated per pulse are ~0.1, ~0.01 and ~0.05 for the first, second and third pair, respectively.

**Preparing spin-orbit entanglement:**

As illustrated in the yellow panel of Fig. 2, we use a Sagnac interferometer to prepare the spin-orbit hybrid entangled state $|\varphi\rangle_{\mathrm{E}}=(|0\rangle^s|0\rangle^o+|1\rangle^s|1\rangle^o)/\sqrt{2}$ to be teleported. The photon 1 is initially prepared in the $(|0\rangle^s-|1\rangle^s)/\sqrt{2}$ SAM state with zero-order OAM. It is then sent into a Sagnac interferometer that consists of a PBS and an SPP, where the counter-propagating $|0\rangle^s$ and $|1\rangle^s$ SAMs with zero-order OAM pass through the SPP in opposite directions and are converted into $|0\rangle^s|0\rangle^o$ and $|1\rangle^s|1\rangle^o$, respectively. Considering an extra $\pi$ phase shift for the $|1\rangle^s$ SAM from double reflections in the PBS, the final output state is $(|0\rangle^s|0\rangle^o + |1\rangle^s|1\rangle^o)/\sqrt{2}$. We note that the whole conversion process is deterministic.

**Dual-channel and efficient OAM measurement:**

One of the most frequently used OAM measurement devices in the previous experiments is off-axis hologram gratings[21,23,24,31], which however typically have a practical efficiency ($p$) of about 30%[31]. This low efficiency would cause an extremely low six-photon coincidence count rate that scales as $p^6$, more challenging than the scaling ($p^2$) in the previous two-photon OAM experiments. To overcome this challenge, we use two different types of devices for efficient OAM readout.

The first type is what we refer to as "dual-channel" OAM measurement devices, used after the two BSs. The strategy is to transfer the OAM information to the photon's SAM, and measure it using a PBS with two output channels. The method is as follows: After the BS, each photon passes through a HWP and is preared in the state $(|0\rangle^s + |1\rangle^s)/\sqrt{2}$. They are then sent into Sagnac interferometers with a Dove prism inside[32] (see Fig. 2), which is placed with a $\pi/8$ angle with respect to the interferometer plane. The photon is rotated by an $\pi/4$ ($-\pi/4$) angle for the $|0\rangle^s$ ($|1\rangle^s$) SAM component when forward (backward) passing through the Dove prism. Considering an OAM state $|0\rangle^o$ ($|1\rangle^o$) with a phase $e^{i\varphi}$ ($e^{-i\varphi}$), where $\varphi$ is the azimuth angle in the polar system passes through a Dove prism rotated at $\pi/8$, it will be rotated by an angle of $\pi/4$ and the phase will be changed to $e^{i(\varphi-\pi/4)}$ [$e^{-i(\varphi-\pi/4)}$]. Therefore, two opposite phase $e^{i\pi/4}$ and $e^{-i\pi/4}$ will be added to the two orthogonal OAM modes. Finally, the output photon is transformed into the $|0\rangle^s$ ($|1\rangle^s$) polarization using a QWP. The overall transformation can be summerised as a CNOT gate between the OAM and SAM:

$$|0\rangle^s|0\rangle^o \xrightarrow{\text{HWP}} (|0\rangle^s + |1\rangle^s)|0\rangle^o/\sqrt{2} \xrightarrow{\text{Sagnac}} e^{i\pi/4}(|0\rangle^s + i|1\rangle^s)|0\rangle^o/\sqrt{2} \xrightarrow{\text{QWP}} e^{i\pi/4}|0\rangle^s|0\rangle^o,$$
$$|0\rangle^s|1\rangle^o \xrightarrow{\text{HWP}} (|0\rangle^s + |1\rangle^s)|1\rangle^o/\sqrt{2} \xrightarrow{\text{Sagnac}} e^{-i\pi/4}(|0\rangle^s - i|1\rangle^s)|1\rangle^o/\sqrt{2} \xrightarrow{\text{QWP}} e^{-i\pi/4}|1\rangle^s|1\rangle^o$$

Effectively, the OAM qubit is deterministically and redudantly encoded by the SAM qubit, i.e., the SAM becomes idential to the OAM state. Thus, by measuring the SAM using a PBS with two output channel, we can derive the informaiton of the OAM with a high efficiency of ~97%. In our experiment, four such Sagnac interferometers are utilized in the *h*-BSM. In this way, two out of the four OAM Bell states can be discriminated using a BS.

While the first type is like a two-channel readout device (like a PBS for SAM), the second type is like a one-channel readout device (like a polarizer for SAM), and is used in the final stage of state verification after teleportation. The strategy for the projective OAM measurement is by transforming it into the zero-order OAM mode so that the photon can be coupled into a single-mode fibre, while all other higher-order OAM modes will be rejected. Here we use SPP[33] and BPP[33] for efficient OAM readout.

SPP[33] is designed with a spiral shape to create a vortex phase of $e^{il\varphi}$, where $l$ is an integer and referred to as topological charge. SPP can attach a vortex phase $e^{il\varphi}$ ($e^{-il\varphi}$) to a photon when it forwardly (backwardly) passes through the SPP. When attached a vortex phase $e^{il\varphi}$ ($e^{-il\varphi}$), the corresponding OAM value will add (subtrate) $l\hbar$. Therefore, a SPP can be used as an OAM mode converter. In our experiment, the OAM qubits are encoded in the OAM first-order subspace with the topological charge $l$ being 1. The conversion between the OAM zero mode and the first-order modes ($|0\rangle^o, |1\rangle^o$) is realized using a 16-phase level SPP with an efficiency of ~97%.

For the coherent transformation between the OAM zero mode and the superposition states $(|0\rangle^o \pm |1\rangle^o)/\sqrt{2}$ and $(|0\rangle^o \pm i|1\rangle^o)/\sqrt{2}$, a 2-phase level BPP[33,34] is used with an efficiency of ~80%. These high-efficiency OAM measurement devices boost the six-fold coincidence count

rate in our present experiment by more than two orders of magnitudes, compared to the previous use of hologram gratings.

**Error budget**

The sources of error of our experiment include: 1. Double pair emission in spontaneous parametric down conversion; 2. Partial distinguishability of the independent photons interfered on the PBS (~5%) and BSs (~5%); 3. State measurement error due to zero-order OAM leakage (~2%); 4. Fidelity imperfection of the entangled photons 2-3 (~5%) and 4-5 (~9%); 5. Fidelity imperfection of the to-be-teleported single-photon hybrid entangled state (~8%).

Some error sources affect all the teleported states. First, the double pair emission contributed ~15% background to the overall six-fold coincidence counts. If this is subtracted, the average teleportation fidelity would be improved to ~0.74. Second, the imperfectly entangled photon pair 4-5 and the imperfect two-photon interference at the $BS_1$ and $BS_2$ (for QND and OAM Bell-state measurement, respectively) degrades the teleportation fidelity for all states by ~13%. Third, the imperfect state measurements mainly due to the zero-order OAM leakage cause another degradation of ~2%.

We analyse that some error sources can affect differently on different teleported states. The imperfect two-photon interference at the PBS degrades the teleportation fidelity for the states $|\varphi\rangle_C$, $|\varphi\rangle_D$, and $|\varphi\rangle_E$ by ~5%. However, for the states $|\varphi\rangle_A$ and $|\varphi\rangle_B$, as the SAM qubits there is in the preferred (*H*/*V*) basis for the PBS, they are immune to the imperfect interference at the PBS. This explains why the teleportation fidelities for the states $|\varphi\rangle_A$ and $|\varphi\rangle_B$ are the highest. This is in consistent with the observation of the previous results: for the experiments using a PBS[35-37], the teleportation fidelities in the *H*/*V* basis is higher than those in the (*H*+*V*)/(*H*-*V*) and (*H*+*iV*)/(*H*-*iV*) basis); whereas for the experiments[38,39] using a non-polarising BS, the teleportation fidelities in all polarization basis are largely unbiased.

We note that the teleportation fidelity for the hybrid entangled state $|\varphi\rangle_{\mathrm{E}}$ is the lowest, which is affected by, on top of those in $|\varphi\rangle_{\mathrm{C}}$ and $|\varphi\rangle_{\mathrm{D}}$, further by the imperfection (~8%) of state preparation of the initial state $|\varphi\rangle_{\mathrm{E}}$ — essentially all error and noise present in the experiment. It can be expected that all entangled states are subject to the same decoherence mechanism as the state $|\varphi\rangle_{\mathrm{E}}$ as demonstrated, and should have similar fidelity reduction.

These noise can be in principle eliminated using various methods in the future experiments. For instance, deterministic entangled photons[40] can avoid the problem of double pair emission. We also plan to develop bright OAM-entangled photons with higher fidelity, and more precise 32-phase level SPP for the next experiment of hyper-entanglement swapping.

**A universal scheme for teleporting *N* DoFs**

We illustrate in Extended Data Figure 2 a universal scheme for teleporting *N* DoFs of a single photon. For simplicity, we discuss an example for three DoFs (Extended Data Figure 2**b**), labelled as X, Y, and Z. There are in total 64 hyper-entangled Bell states: $\{|\phi\rangle_{\mathrm{X}}^{+},|\phi\rangle_{\mathrm{X}}^{-},|\psi\rangle_{\mathrm{X}}^{+},|\psi\rangle_{\mathrm{X}}^{-}\}\otimes\{|\phi\rangle_{\mathrm{Y}}^{+},|\phi\rangle_{\mathrm{Y}}^{-},|\psi\rangle_{\mathrm{Y}}^{+},|\psi\rangle_{\mathrm{Y}}^{-}\}\otimes\{|\phi\rangle_{\mathrm{Z}}^{+},|\phi\rangle_{\mathrm{Z}}^{-},|\psi\rangle_{\mathrm{Z}}^{+},|\psi\rangle_{\mathrm{Z}}^{-}\}$, which are the products of the Bell states of each DoF, as defined below:

$|\phi\rangle_i^{\pm}=(|0\rangle_i|0\rangle_i\pm|1\rangle_i|1\rangle_i)/\sqrt{2}$ and $|\psi\rangle_i^{\pm}=(|0\rangle_i|1\rangle_i\pm|1\rangle_i|0\rangle_i)/\sqrt{2}$, where $i$ = X, Y, Z.

The aim is to perform an *h*-BSM, identifying 1 out of the 64. The required resources include: photon pairs entangled in the Z DoF, hyper-entangled in the Y-Z DoFs, and hyper-entangled in the X-Y-Z DoFs, filters for the three DoFs that can project the state to $|0\rangle$ or $|1\rangle$ (with a functionality similar to the polarizers for SAM), qubit flip ($\sigma_{\mathrm{x}}$) operations for the DoFs, 50:50 non-polarising BS, and single-photon detectors, all of which are commercially available or have been experimentally demonstrated previously.

It has been known that if two single photons are superposed on a BS, only asymmetric quantum states can lead to the case that one and only one photon exits from each output of the BS. In the simplest case of one DoF, this is the asymmetric $|\psi\rangle^-$ state. As now we have three DoFs, we have to consider the combined, molecular-like quantum states. There are in total 28 possible combination that are asymmetric states, as listed below:

$|\psi\rangle_X^- \otimes \{|\phi\rangle_Y^+, |\phi\rangle_Y^-, |\psi\rangle_Y^+\} \otimes \{|\phi\rangle_Z^+, |\phi\rangle_Z^-, |\psi\rangle_Z^+\}$ , $\{|\phi\rangle_X^+, |\phi\rangle_X^-, |\psi\rangle_X^+\} \otimes |\psi\rangle_Y^- \otimes \{|\phi\rangle_Z^+, |\phi\rangle_Z^-, |\psi\rangle_Z^+\}$ ,

$\{|\phi\rangle_X^+, |\phi\rangle_X^-, |\psi\rangle_X^+\} \otimes \{|\phi\rangle_Y^+, |\phi\rangle_Y^-, |\psi\rangle_Y^+\} \otimes |\psi\rangle_Z^-$,

$|\psi\rangle_X^- \otimes |\psi\rangle_Y^- \otimes |\psi\rangle_Z^-$.

After passing through the first BS, we can filter out these 28 from the total 64 hyper-entangled Bell states conditioned on seeing one and only one photon in each output of the BS. Next, we apply two filters in the two output of the BS to project the X DoF into $|\psi\rangle^\pm$. One filter is set to pass the $|1\rangle$ state and the other is set to pass $|0\rangle$ state. This results in the below 16 filtered states from the 28, as listed below:

$|\psi\rangle_X^- \otimes \{|\phi\rangle_Y^+, |\phi\rangle_Y^-, |\psi\rangle_Y^+\} \otimes \{|\phi\rangle_Z^+, |\phi\rangle_Z^-, |\psi\rangle_Z^+\}$,

$|\psi\rangle_X^+ \otimes |\psi\rangle_Y^- \otimes \{|\phi\rangle_Z^+, |\phi\rangle_Z^-, |\psi\rangle_Z^+\}$,

$|\psi\rangle_X^+ \otimes \{|\phi\rangle_Y^+, |\phi\rangle_Y^-, |\psi\rangle_Y^+\} \otimes |\psi\rangle_Z^-$,

$|\psi\rangle_X^- \otimes |\psi\rangle_Y^- \otimes |\psi\rangle_Z^-$.

We perform a bit-flip operation on the X DoF on one of the arms, erasing the information on the X DoF. We then pass them into the second BS, filtering out the asymmetric 6 combinations:

$|\psi\rangle_X^+ \otimes \{|\psi\rangle_Y^-\} \otimes \{|\phi\rangle_Z^+, |\phi\rangle_Z^-, |\psi\rangle_Z^+\}$,

$|\psi\rangle_X^+ \otimes \{|\phi\rangle_Y^+, |\phi\rangle_Y^-, |\psi\rangle_Y^+\} \otimes |\psi\rangle_Z^-$.

We emphasize that, before sending the photons to the seond BS, due to the same reason discussed in the main text, we use teleportation-based QND measurement to ensure the two photons can be further fed into the following cascaded interferometers. Here, the QND should preserve the quantum information in the Y and Z DoF. Thus, a quantum teleportation of two DoFs of a single photons is required as a key module (Extended Data Figure 2**a**), which is exactly what we have demonstrated in the experiment presented in the main text.

After that, we again pass them through two filters on the Y DoF, one set in the $|1\rangle$ and the other one set in the $|0\rangle$ state, which leaves 4 asymmetric combinations:

$|\psi\rangle_X^+ \otimes \{|\psi\rangle_Y^-\} \otimes \{|\phi\rangle_Z^+, |\phi\rangle_Z^-, |\psi\rangle_Z^+\}$, and $|\psi\rangle_X^+ \otimes |\psi\rangle_Y^+ \otimes |\psi\rangle_Z^-$.

Similarly, we perform a bit-flip operation on the Y DoF on one of the arms, erasing the information on the Y DoF. We then do a QND measurement on the Z DoF, and pass the two photons into the third BS, finally filtering out the only remaining asymmetric state:

$|\psi\rangle_X^+ \otimes |\psi\rangle_Y^+ \otimes |\psi\rangle_Z^-$.

By detecting one and only one photon out of the third BS, we can thus discriminate this particular hyper-entangled state $|\psi\rangle_X^+ \otimes |\psi\rangle_Y^+ \otimes |\psi\rangle_Z^-$ from the 64 hyper-entangled Bell states on three DoFs.

To experimentally demonstrate the teleportation of three DoFs, the scheme would need in total 10 photons (or 5 entangled photon pairs from SPDC), which is within the reach of near future experimental abilities given the recent advances in high-efficiency photon collection and detection. It is obvious that the above protocol can be extended to more DoFs as displayed in Extended Data Figure 2**c**.

**Feed-forward scheme for spin-orbit composite states**

To realize a deterministic teleportation, feed-forward Pauli operations on the teleported particle based on the intrinsically random Bell-state measurement results are essential. In our present experiment, no feed-forward has been applied. Here we briefly describe how this can be done for the spin-orbit composite state. For the SAM qubits, active feed-forward has been demonstrated before using fast electro-optical modulators (EOMs)[39,41]. To take advantage of this existing technology with demonstrated high-speed operation and high gate fidelity, we employ a coherent quantum **SWAP** gate between the OAM and SAM qubits. The **SWAP** gate is defined as:

$$\begin{aligned}&(\alpha|0\rangle_S|0\rangle_O+\beta|1\rangle_S|0\rangle_O+\gamma|0\rangle_S|1\rangle_O+\delta|1\rangle_S|1\rangle_O)\\ &\xrightarrow{\text{SWAP}}(\alpha|0\rangle_O|0\rangle_S+\beta|1\rangle_O|0\rangle_S+\gamma|0\rangle_O|1\rangle_S+\delta|1\rangle_O|1\rangle_S)\end{aligned}.$$

The operation sequence for the feed-forward operation on the spin-orbit composite states is shown in the Extended Data Figure 3a. First, the SAM feed-forward is done with an EOM. Second, the SAM and OAM qubits undergo a **SWAP** gate. Third, an EOM is used to operate on the "new" SAM that is converted from the OAM. Note that the EOM doesn't affect the OAM, so that the previous operation on the SAM is unaffected. Lastly, a final **SWAP** gate convert the SAM back to OAM, which completes the feed-forward for spin-orbit composite states.

The **SWAP** gate is composed of three CNOT gates (see the Extended Data Fig.3b). In the first and third CNOT gate (blue shading), the SAM qubit is the control qubit that acts on the OAM as target qubit, realized by sending a photon through, and recombined on a PBS. On the reflection arm only, a Dove prism is inserted to induce an OAM bit flip. In the second CNOT (yellow shading), the control (target) qubit is the SAM (OAM), as explained in the Methods section: "Dual-channel and efficient OAM measurement". The extra Dove prism and half-wave plate in front of the PBS are used to compensate for the phase shift inside the Sagnac interferometer.

## References:

31. Graham, T. M. Barreiro, J. T. Mohseni, M & Kwiat, P. G, Hyperentanglement-enabled direct characterization of quantum dynamics. *Phys. Rev. Lett.* **110**, 060404 (2013).
32. Slussarenko, S. *et al*. The polarizing Sagnac interferometer: a tool for light orbital angular momentum sorting and spin-orbit photon processing. *Opt. Express* **18**, 27205–27216 (2010).

33. Andrews, D. L. & Babiker, M. (ed.) *The Angular Momentum of Light* (Cambridge Univ. Press, Cambridge, UK, 2012).

34. Jack, B., Leach, J., Ritsch, H., Barnett, S. M., Padgett, M. J. & Franke-Arnold, S. Precise quantum tomography of photon pairs with entangled orbital angular momentum. *New J. Phys.* **11** 103024 (2009).

35. Pan, J.-W., Gasparoni, S., Aspelmeyer, M., Jennewein, T. & Zeilinger, A. Experimental realization of freely propagating teleported qubits. *Nature* **421**,721–725 (2003).

36. Zhang, Q. *et al.* Experimental quantum teleportation of a two-qubit composite system. *Nature Phys.* **2**, 678–682 (2006).

37. Yin, J. *et al*. Quantum teleportation and entanglement distribution over 100-kilometre free-space channels. *Nature* **488**, 185–188 (2012).

38. Bouwmeester, D. *et al.* Experimental quantum teleportation. *Nature* **390**, 575–579 (1997).

39. Ma, X.-S. *et al*. Quantum teleportation over 143 kilometres using active feed-forward. *Nature* **489**, 269–273 (2012).

40. Lu, C.-Y., & Pan, J.-W., Push-button photon entanglement. *Nature Photonics* **8**, 174–176 (2014).

41. Prevedel, R. *et al*. High-speed linear optics quantum computing using active feed-forward. *Nature* **445**, 65–69 (2007).

## Extended Data Figure captions

**Extended Data Figure 1 | Hong-Ou-Mandel interference of multiple independent photons encoded with SAM or OAM**. **a**, Interference on the PBS where two input photons 1 and 2 are intentionally prepared in the states $D_1 r_1 A_2 l_2$ (orthogonal SAMs, see open squares) and $D_1 r_1 D_2 l_2$ (parallel SAMs, see solid circles). We measure each single photon from the two outputs of the PBS in the $D_{1'} D_{2'}$ basis (1' and 2' denote the output spatial modes). At zero delay where the photons are optimally overlapped in time, the orthogonal SAM input yields an output state of $(|0\rangle_{1'}^{s}|0\rangle_{2'}^{s} + |1\rangle_{1'}^{s}|1\rangle_{2'}^{s})/\sqrt{2}$ conditioned on a coincidence detection, which can be decomposed in the diagonal basis as $(|D\rangle_{1'}^{s}|\mathrm{D}\rangle_{2'}^{s} + |A\rangle_{1'}^{s}|A\rangle_{2'}^{s})/\sqrt{2}$, thus showing an enhancement. For the parallel SAM input, the output stat is $(|D\rangle_{1'}^{s}|A\rangle_{2'}^{s} + |A\rangle_{1'}^{s}|D\rangle_{2'}^{s})/\sqrt{2}$, which shows a dip. The increase of the delay gradually destroys the indistinguishability of the two photons so their quantum state becomes a classical mixture. Thus, at large delays the counts appear flat. This type of interferometer is sensitive only to length changes on the order of the coherence length of the detected photons and stay stable for weeks. The y axis shows the raw four-fold (the trigger photon *t*, photon 1, 2 and 3) coincidence counts. The extracted visibility is $0.75 \pm 0.03$, calculated by $\mathrm{V}(0) = (\mathrm{C}_{+} - \mathrm{C}_{//})/(\mathrm{C}_{+} + \mathrm{C}_{//})$, where $\mathrm{C}_{//}$ ($\mathrm{C}_{+}$) is the coincidence counts without any background subtraction at zero delay for parallel (orthogonal) SAM, respectively. The red (blue) line is a Gaussian fit to the raw data. **b**, Two-photon interference on the $\mathrm{BS}_1$ for the teleportation-based QND measurement of OAM qubits. Two input photons 1 and 4 are prepared in the same SAM but orthogonal OAM states. It is interesting to note that, in stark contrast to the conventional Hong-Ou-Mandel interference, only when the two input OAM states are orthogonal can lead to an interference dip, because the reflection on the BS flips the sign of OAM. The black line is a Gaussian fit to the raw data of fourfold (the trigger photon, photon 1, 4 and 5) coincidence counts. The visibility yields $0.73 \pm 0.03$, calculated by $\mathrm{V}(0) = 1\text{-}\mathrm{C}_0/\mathrm{C}_{\infty}$,

where $C_0$ ($C_\infty$) is the fitted counts at zero (infinite) delay. **c**, Two-photon interference on the $BS_2$ for BSM of OAM qubits. Two input photons 1 and 5 are in intentionally prepared in the orthogonal OAM states. The black line is a Gaussian fit to the data points. The interference visibility yields $0.69 \pm 0.03$ calculated in the same way as in **b**.

**Extended Data Figure 2 | A universal scheme for teleporting *N* DoFs of a single photons**. **a**, A schem for quantum teleportation of two DoFs of a single photons using three BSs, which is slightly different from the scheme presented in the main text using a PBS and two BSs. Through the first BS, the 6 states, $|\psi\rangle_X^- \otimes \{|\phi\rangle_Y^+, |\phi\rangle_Y^-, |\psi\rangle_Y^+\}$ and $\{|\phi\rangle_X^+, |\phi\rangle_X^-, |\psi\rangle_X^+\} \otimes |\psi\rangle_Y^-$, that are asymmetric will result in one and only one photon in each output. Teleportation-based QND on the Y DoF is used to ensure the presence of one single photon in each output. After pssing the two photos through the two filters that project them in the $|1\rangle$ and $|0\rangle$ state on the X DoF, 4 states, $|\psi\rangle_X^- \otimes \{|\phi\rangle_Y^+, |\phi\rangle_Y^-, |\psi\rangle_Y^+\}$ and $|\psi\rangle_X^+ \otimes |\psi\rangle_Y^-$, survive. Sending them through the second BS, only the asymmetric state $|\psi\rangle_Y^-$ on Y DoF will result in one and only one in each output. Thus we can discriminate the state $|\psi\rangle_X^+ \otimes |\psi\rangle_Y^-$ from the 16 hyper-entangled Bell state. **b**, Quantum teleportation of three DoFs of a single photons. Based on the conditional detection of one and only one photon in each output of BS, the first BS will post-select 28 asymmetric states from the total 64 hyper-entangled Bell state in three DoFs X Y and Z, then two filters in X DoF will futher post-select 16 states from the 28 states. Note that the condition that one and only one photon in the output of the first BS is fulfilled by the teleportation-based QND in two DoFs in **a**. Similarliy, the second BS will post-select 6 states from the remaining 16 and two filters in Y DoF will post-select 4 state from the 6 states. Finally the third BS will post-selcet 1 state from the last 4 states. Then a *h*-BSM in three DoFs will be accomplished, which is necessary for the quantum teleportation of three DoFs of a single photons. See Methods for details. **c**, Quantum teleportation of *N* DoFs of a single photons. The most crucial step is a *h*-BSM in *N* DoFs, which

could be implemented based on the three principles: (1) BS could post-select the asymmetric hyper-entangled Bell states in $N$ DoFs which contain odd numer of asymmetric Bell state in one DoF, (2) two filters and one bit flip operation will erase the information on the measured DoF and further post-select asymmetric states, (3) teleportation-based QND in less DoFs will enable one and only one photon pass through the next BS, which will allow the connection of the cascaded interferometers.

**Extended Data Figure 3 | Active feed-forward for spin-orbit composite states**. **a**, The active feed-forward scheme. This composite active feed-forward could be completed in a step-by-step manner. First, we use a EOM to implement the active feed-forward for SAM qubits. Second we employ a coherent quantum **SWAP** gate between the OAM and SAM qubits. The orignal OAM will be converted into the "new" SAM, which active feed-foward will be completed by a second EOM. Then the OAM and SAM qubits will undergo a second **SWAP** gate and be converted to the orginal DoFs. Because EOM doesn't affect OAM, the first EOM completes the active feed-forward for the original SAM and the second EOM is utilized for the "new" SAM corresponding the orginal OAM. Therefore, two EOMs and two **SWAP** gates will enable the active feed-forward for spin-orbit composite states. **b**, The quantum circuit for a **SWAP** gate between the OAM and SAM qubits. The **SWAP** gate is composed of three CNOT gates: in the first and the third CNOT gates, the SAM qubit acts as the control qubit and the OAM qubit is the target qubit, while in the second CNOT gate, the OAM (SAM) acts as the control (target) qubit.

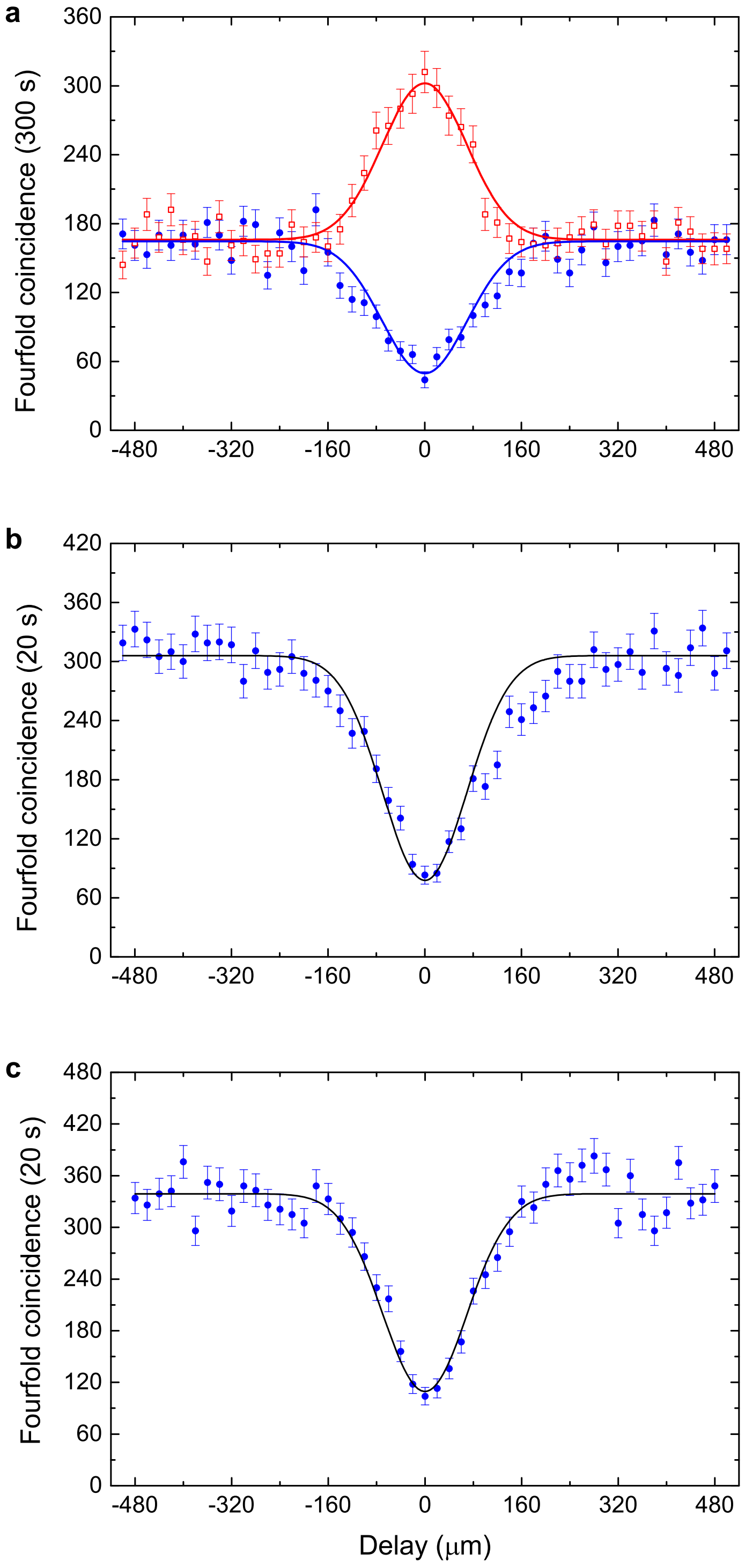

Extended Data Figure 1

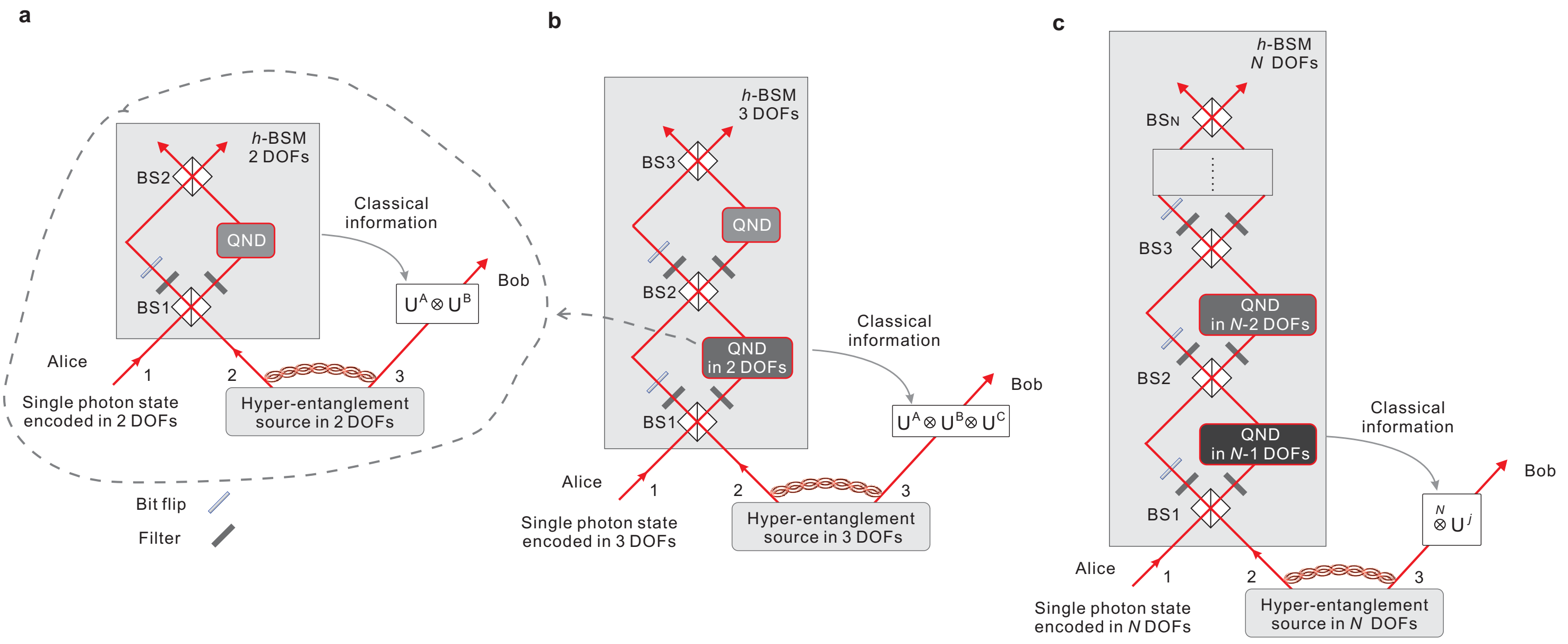

Extended Data Figure 2

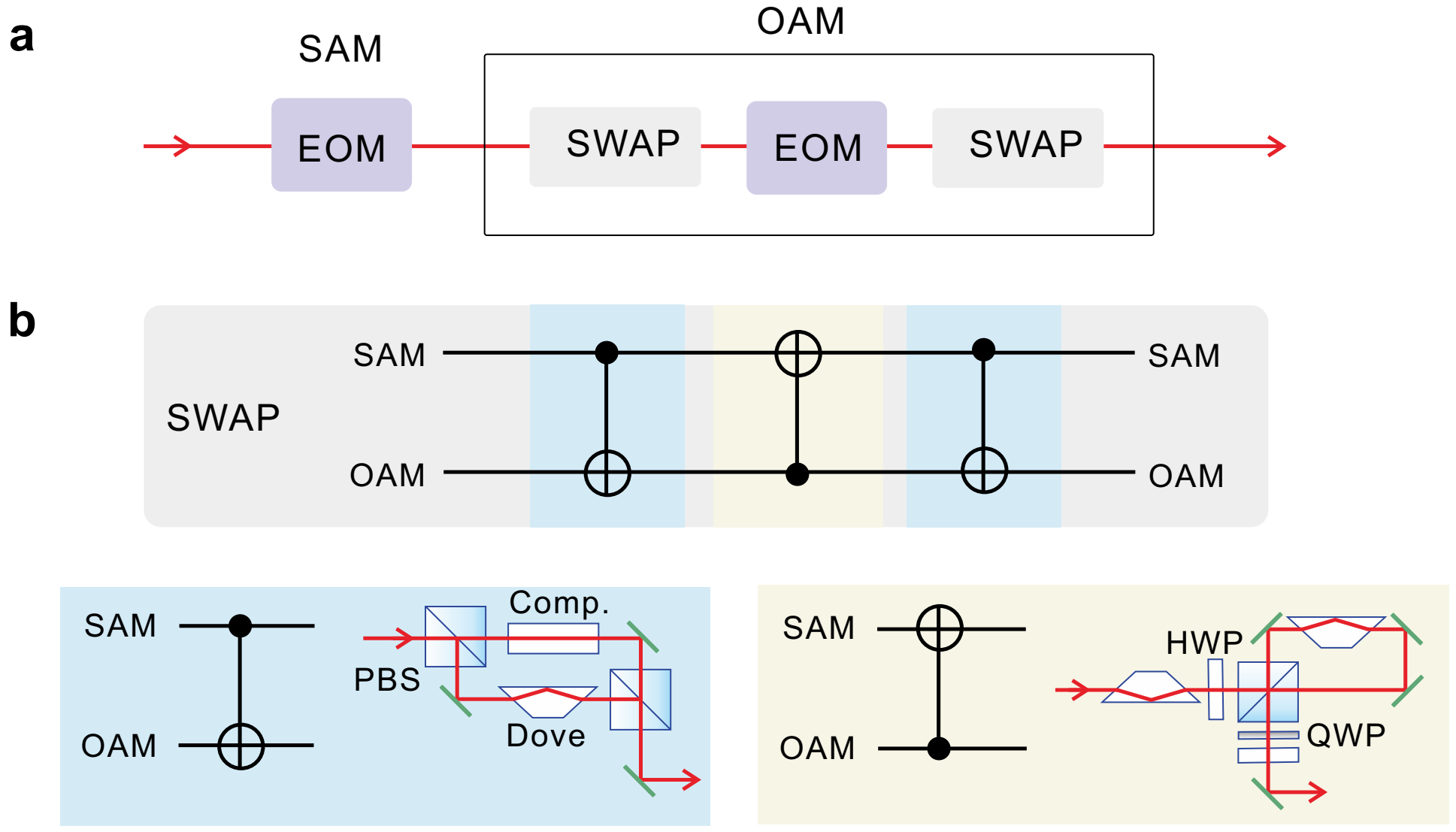

Extended Data Figure 3